\documentclass[11pt,a4paper]{article}
\usepackage{jcappub}
\usepackage{amssymb,graphicx}
\usepackage{epstopdf}
\usepackage{amsmath,amsfonts}
\usepackage{epsfig}

\def\d{\partial}
\def\l{\left(}
\def\r{\right)}

\newcommand{\be}{\begin{equation}}
\newcommand{\ee}{\end{equation}}
\newcommand{\bea}{\begin{eqnarray}}
\newcommand{\eea}{\end{eqnarray}}
\newcommand{\bg}{\begin{gather}}
\newcommand{\eg}{\end{gather}}
\newcommand{\bseq}{\begin{subequations}}
\newcommand{\eseq}{\end{subequations}}

\title{ Cosmological density perturbations\\%[0.3cm]
from conformal scalar field:
infrared properties and statistical anisotropy}
\author[]{M.~Libanov,}
\author[]{V.~Rubakov}
\affiliation{
Institute for Nuclear Research of
         the Russian Academy of Sciences,\\  60th October Anniversary
  Prospect, 7a, 117312 Moscow, Russia}
\emailAdd{ml@ms2.inr.ac.ru}
\emailAdd{rubakov@ms2.inr.ac.ru}

\abstract{
We consider a scenario in which primordial scalar perturbations
are generated when complex conformal scalar field rolls down its
negative quartic potential. Initially, these are the perturbations of the
phase of this field; they are converted into the adiabatic perturbations at
a later stage. A potentially dangerous feature of this scenario
is the existence of perturbations in the radial field direction, which have
red  power spectrum. We show, however, that
to the linear order in the small parameter --- the quartic self-coupling
--- the infrared effects are completely harmless, as they can be absorbed
into field redefinition. We then evaluate the statistical anisotropy
inherent in the model due to the existence of the long-ranged radial
perturbations. To the linear order in  the quartic self-coupling the
statistical anisotropy is free of the infrared effects. The latter show up
at the quadratic order in the self-coupling and result in the mild
(logarithmic) enhancement of the corresponding contribution to the
statistical anisotropy. The resulting statistical anisotropy is a
combination of a larger  term which, however,
decays as momentum increases, and a smaller
term which is independent of momentum.
}
\arxivnumber{1007.4949}
\begin{document}
\maketitle
\flushbottom
\section{Introduction and summary}

The two basic properties of
primordial scalar perturbations in the Universe are
approximate Gaussianity and approximate flatness of the power
spectrum~\cite{Komatsu:2010fb}. The first property strongly
suggests that these perturbations
originate from amplified vacuum fluctuations of nearly
linear (i.e., weakly coupled) quantum field(s): free quantum
field in its vacuum state
obeys the Wick theorem, the defining property of
Gaussian random field, while linear evolution in classical backgrounds
does not induce the non-Gaussianity.
% into the field fluctuations.
The flatness of the power spectrum calls for some symmetry behind
it.
The best known candidate
is the symmetry of the de~Sitter space-time under
 spatial dilatations supplemented by time translations.
This is the approximate
symmetry of the inflating Universe~\cite{inflation}, and, indeed,
the inflationary mechanism of the generation of scalar
perturbations~\cite{infl-perturbations} produces almost flat
power spectrum. Another example is the symmetry (of the field equation)
of the scalar theory with negative exponential scalar potential
in flat space-time. This is the symmetry
 under space-time dilatations suppelemented by
the shifts of the field. This symmetry remains approximately valid in
slowly evolving, e.g., ekpyrotic~\cite{ekpyrosis} or
``starting''~\cite{starting}
Universe, and the resulting perturbation
spectrum is again almost
flat\footnote{There are other mechanisms capable of producing flat or
almost flat scalar
spectrum~\cite{Wands:1998yp,Mukohyama:2009gg,Creminelli:2010ba}.
In some cases, there is no obvious symmetry
that guarantees the flatness, i.e., the scalar spectrum is
 flat accidentally.}~\cite{minus-exp} (see also ref.~\cite{minus-old}).
%\marginpar{\bf Matter bounce; Mukohyama; recent Creminelli}

Yet another symmetry that may be responsible for the approximate
flatness of the scalar spectrum has been proposed  in
ref.~\cite{vrscalinv}. This is a combination of conformal invariance and
a global symmetry. The simplest model involves complex scalar field
$\phi$, which is conformally coupled to gravity and for long enough time
evolves in
negative quartic potential
\be
V(\phi) = - h^2 |\phi|^4 \; .
\label{jul22-1}
\ee
A necessary condition for the absence of strong non-linearities
at the classical level and strong coupling at the quantum level is
\[
 h < 1 \; .
\]
As discussed in section~\ref{Reprocess}, there may or may not be
stronger bounds on $h$.
The global symmetry in this simplest case is $U(1)$ acting as
$\phi \to \mbox{e}^{i\alpha} \phi$.
One assumes that the background space-time is homogeneous,
isotropic and spatially flat,
$ds^2 = a^2(\eta)(d\eta^2 - d{\bf x}^2)$. Then, due to conformal
invariance, the dynamics of the field
\[
\chi = a \phi
\]
 is independent of the evolution of
the scale factor and in terms of the conformal coordinates
$(\eta, {\bf x})$ proceeds in the same way as in Minkowski space-time.
One begins with the homogeneous background field $\chi_c (\eta)$ that
rolls down the negative quartic potential. Its late-time behavior
is completely determined by conformal invariance.
As we review in section~\ref{Review}, this
rolling field produces effective ``horizon'' for perturbations
$\delta \chi$: at early times the linear
perturbations oscillate in conformal time
as modes of
free massless (quantum) scalar field,\footnote{An assumption here is that
the rolling stage is long enough in conformal time, so that the modes of
interest are indeed sub-''horizon'' at early times. This
%e latter
%condition
assumption is non-trivial: the conformal rolling should occur at a
cosmological epoch, preceeding the hot Big Bang stage,
 when the standard horizon problem
is solved, at least formally, see ref.~\cite{vrscalinv} and
section~\ref{Review}.} while at late times the oscillatory
behavior ceases to hold. The perturbations of the phase $\theta
=\mbox{Arg}~\phi$
freeze out at the time of the ``horizon'' exit, and their power spectrum
is flat after that,
\be
  \sqrt{{\cal P}_{\delta \theta}} = \frac{h}{2\pi} \; .
\label{jul21-1}
\ee
As discussed in ref.~\cite{vrscalinv}, this property is a consequence
of conformal and global symmetries.

\begin{figure}[tb!]
\begin{center}
\includegraphics[width=0.7\textwidth,angle=0]{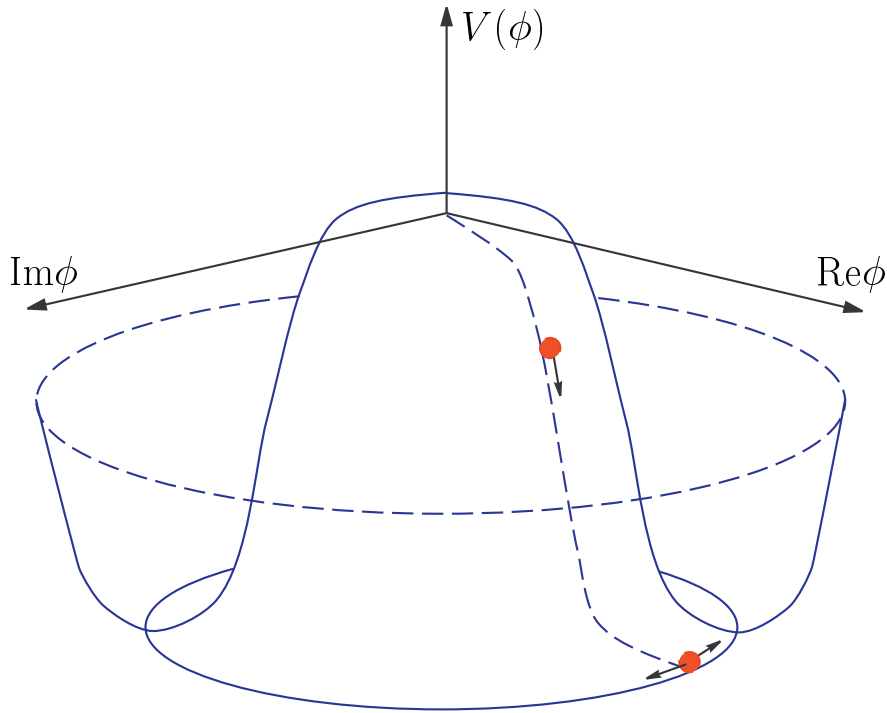}
\end{center}
\caption{The scalar potential. Bullets show the evolution of
the scalar field. Arrows at the end point
at the bottom of the potential indicate that there are perturbations of
the phase.
\label{fig1}}
\end{figure}
The scenario proceeds with the assumption that the scalar potential
$V(\phi)$ has, in fact, a minimum at some large value of $|\phi|$,
and that the modulus of the field $\phi$ eventually gets relaxed
to the minimum, see figure~\ref{fig1}.
The simplest option  concerning further evolution of the
perturbations $\delta \theta$ is that they are superhorizon in the
conventional sense by the time the conformal rolling stage
ends.  Then they remain frozen out,\footnote{For contracting Universe,
this property of superhorizon modes holds if the dominating matter
has stiff equation of state, $w>1$. This appears to be necessary for the
viability of the bounce scenario anyway, see the discussion in
refs.~\cite{smooth, ekpyro}.} and their power spectrum remains flat.
At some much later cosmological epoch, the perturbations of the phase
are converted into the adiabatic scalar perturbations; we discuss possible
mechanisms responsible for that in section~\ref{Reprocess}.
These mechansims do not distort the power spectrum, so the resulting
adiabatic perturbations have flat primordial power spectrum.
If conformal invariance is not exact at the rolling stage,
the scalar power spectrum has small tilt, which depends on both
the strength of the violation of conformal invariance and the
evolution of the scale factor at the rolling stage~\cite{Osipov}.

A peculiar, and potentially dangerous
 property of the model is that the modulus of the
rolling field also acquires perturbations. Super-''horizon'' modes
of the modulus (i.e., radial direction)
have red power spectrum (see section~\ref{modulusperturbations}
for details),
\be
\sqrt{{\cal P}_{|\phi|}(k)} \propto k^{-1} \; .
\label{jul23-2}
\ee
One consequence is that there are perturbations of the energy density
with red spectrum
right after the conformal rolling stage, but before the modulus freezes out
at the minimum of $V(\phi)$. These are not dangerous, provided that
the energy density of the field $\phi$ is small compared to the total
energy density at all times preceding the time when
the modulus of this field settles down
to the minimum of its potential, i.e., the cosmological evolution is
governed by some other matter at that early epoch.
In this paper we assume that this is indeed the case.

The second consequence is that the infrared radial modes
interact with
the perturbations of the phase, and in principle may have strong effect on
the latter. This is precisely the issue we address
in this paper. To this end, we study the perturbations of the phase
in the presence of long-ranged perturbations of the modulus and make use
of the spatial gradient expansion of the latter. We interpret the effects
emerging at the zeroth  and first orders in the gradient expansion as
the local time shift and local Lorentz boost of the background
field $\chi_c$, the latter property meaning  that the rolling
background is approximately
homogeneous in a local reference frame different from
the cosmic frame where the metric has the standard
Friedmann-Robertson-Walker form.
Our interpretation makes it straightforward to
obtain the expression for the perturbation
of the phase, valid to these two orders of the gradient expansion.
We then show that
%{\marginpar {\bf \tiny Changed}}
to the linear order in $h$, the
infrared effects
cancel out: the perturbations
of the phase remain Gaussian random and have flat spectrum
\eqref{jul21-1}. Since the %infrared modes contribute to the
red power spectrum is characteristic of
$\delta |\phi|$ and $\d_i (\delta |\phi|)$, but not of
higher spatial derivatives of $\delta |\phi|$,
%this cancellation, occuring
%to the leading
%and subleading orders in the gradient expansion,
our analysis is sufficient to show that,
in fact, there is no gross modification of the results of the linear
analysis due to the
effect of the infrared modes.

The large wavelength modes of $\delta |\phi|$ are not entirely
negligible, however.
The modes  whose present wavelengths exceed the
present Hubble size $H_0^{-1}$ induce statistical anisotropy
in the perturbations of the phase $\delta \theta$, and hence in the resulting
adiabatic perturbations.
%{\marginpar {\bf \tiny Changed to end of next-to-next paragraph}}
To the linear order in $h$, the statistical anisotropy
is generated at the second
%This effect occurs  at the sub-subleading
order in the
gradient expansion, and hence it is free of the infrared effects.
The infrared modes show up at the second order in
$h$, and lead to mild enhancement of the statistical anisotropy at
this order.
%The anisotropy is of the first order in
%the coupling $h$, has general tensorial form and is inversely proportional to
%the spatial momentum,
Accordingly, the power spectrum of the adiabatic
perturbation $\zeta$ has the following form,
\be
{\cal P}_\zeta ({\bf k}) =
{\cal P}_0 (k) \left(1 + c_1 \cdot h \cdot \frac{H_0}{k} \cdot \hat{k}_i
\hat{k}_j w_{ij}
- c_2 \cdot h^2 \cdot ({\bf  \hat k u})^2\right)\; .
\label{jul22-6}
\ee
In the first non-trivial term,
$w_{ij}$ is a traceless symmetric tensor of a general form with unit
normalization, $w_{ij} w_{ij} =1$,
${\bf \hat k}$ is a
unit vector, ${\bf \hat k}= {\bf k}/k$, and $c_1$ is a constant of order
1 whose actual value is undetermined because of the cosmic variance.
As we alluded to above, the deep infrared modes are irrelevant for this term.
In the last term, ${\bf u}$ is some unit vector independent of
$w_{ij}$, and the positive
parameter $c_2$
is logarithmically enhanced due to the infrared effects. This is the first
place where the deep infrared modes show up. Clearly, their effect is
subdominant for small $h$.

The statistical anisotropy encoded in the second term in
\eqref{jul22-6} is similar to
that commonly discussed in inflationary context~\cite{aniso}, and, indeed,
generated in some concrete inflationary
models~\cite{soda}: %(see, however, ref.~\cite{Peloso}):
it does not decay as momentum increases
and has special tensorial form $({\bf \hat{k} u})^2$ with constant
${\bf u}$. On the other hand, the first non-trivial term in
\eqref{jul22-6} has the general tensorial structure and decreases
with momentum. The latter property
%Rather,
%the decrease of the statistical anisotropy with momentum
is somewhat similar to the situation that occurs in
cosmological models with
the anisotropic expansion before inflation~\cite{Peloso}.
Overall, the statistical anisotropy \eqref{jul22-6}
may be quite substantial, since
there are no strong bounds on $h$ at least for one mechanism of conversion of
the phase perturbations into adiabatic ones, as discussed in
section~\ref{Reprocess}.

Interestingly, the overall statistical anisotropy
is a combination of a larger (order O($h$)) term which, however,
%statistical anisotropy
decays as $k$ increases, and a smaller (order O($h^2 \log \Lambda)$, where
$\Lambda$ is the infrared cutoff)
term which is independent of $k$. We consider this feature as a
potential smoking-gun property of our scenario.

It is worth noting that the non-linearity of the scalar potential
gives rise to the non-Gaussianity of the perturbations of the phase
$\delta \theta$, and
hence  the adiabatic perturbations in our scenario, over and beyond
the non-Gaussianity that may be generated at the time
when the phase perturbations
get reprocessed into the adiabatic perturbations. In view of the result
outlined above, this non-Gaussianity is not plagued by the infrared
effects
%{\marginpar {\bf \tiny Change}}
at the first non-trivial order in
$h$, and must be fully calculable at this order.
We do not consider the
non-Gaussianity in this paper,
since
the derivative expansion approach
we employ here is useless in this regard.

The paper is organized as follows. In section~\ref{Review} we review the
linear analysis of the model. In section~\ref{twoorders} we study the
effect of infrared modes of the modulus $\delta |\phi|$
on the perturbations of the phase $\delta \theta$ at the leading and subleading
orders of the gradient expansion,
%{\marginpar {\bf \tiny Change}}
and to the linear order in $h$.
Statistical anisotropy, which is
generated at the next order, is analysed in
section~\ref{anisotropy}. We conclude in section~\ref{conclude}.

\section{Linear analysis}
\label{Review}

At the conformal rolling stage, the dynamics of the scalar field
is governed by the action
\be
S[\phi] =
\int d^4x \sqrt{-g} \left[ g^{\mu\nu}\d_\mu\phi^* \d_\nu\phi
+ \frac{R}{6} \phi^* \phi - V(\phi) \right] \; .
%\label{jul8-10}
\nonumber
\ee
where the scalar potential $V(\phi)$ is negative
and has conformally invariant form \eqref{jul22-1}.
In terms of the field $\chi = a\phi$, the action in conformal coordinates
is the same as in Minkowski space-time,
\be
S [\chi] = \int~d^3x~d\eta~
\left[
\eta^{\mu \nu} \d_\mu \chi^* \d_\nu \chi + h^2 |\chi|^4
\right] \; .
\nonumber
\ee
The field equation is
\be
\eta^{\mu \nu} \d_\mu \d_\nu \chi - 2h^2 |\chi|^2 \chi =0 \; .
\label{jul24-2}
\ee
Spatially homogeneous solutions approach the late-time attractor
\be
\chi_c (\eta) = \frac{1}{h (\eta_* - \eta)} \; ,
\label{jul22-2}
%\nonumber
\ee
where $\eta_*$ is an arbitrary real parameter, and we
consider real solution, without loss of generailty. We take the solutuion
\eqref{jul22-2}
as the background. The meaning
of the parameter $\eta_*$ is that the field $\chi_c$ would run away
to infinity as $\eta \to \eta_*$,
if the scalar potential remained negative quartic at arbitrarily
large  fields.

\subsection{Perturbations of  phase}

At the linearized level, the perturbations of the phase and modulus of the
field $\phi$ decouple from each other. Let us begin with the
perturbations of the phase, or, for real background \eqref{jul22-2},
perturbations of the imaginary part \[
\chi_2 \equiv \frac{\mbox{Im}~ \chi}{\sqrt{2}} \ .
\]
They obey the linearized equation,
%which in momentum
%representation reads
\be
(\delta \chi_2)^{\prime \prime}
%+
%k^2 \delta \chi_2
- \d_i \d_i \; \delta \chi_2
- 2 h^2 \chi_c^2 \; \delta \chi_2  = 0 \; ,
\label{jul25-11}
%\nonumber
\ee
where prime denotes the derivative with respect to $\eta$.
Explicitly,
\be
(\delta \chi_2)^{\prime \prime}
%+ k^2 \delta \chi_2
- \d_i \d_i \; \delta \chi_2
- \frac{2}{(\eta_* - \eta)^2} \; \delta \chi_2  = 0 \; .
\label{jul22-5}
%\nonumber
\ee
Let ${\bf k}$ be conformal momentum of perturbation.
An important assumption of the entire scenario is that the rolling stage
begins early enough, so that there is time at which the following
inequality holds:
\be
   k (\eta_* - \eta) \gg 1 \; .
\label{jul22-3}
\ee
Since the momenta $k$ of cosmological significance are as small
as the present Hubble parameter,
this inequality means that the
duration of the rolling stage in conformal time is longer than the conformal
time elapsed from, say, the beginning of the hot Big Bang expansion to
the present epoch. This is only possible if the hot Big Bang stage
was preceded by some other epoch, at which the standard horizon problem is
solved; the mechanism we discuss in this paper is meant to operate
 at that epoch.
We note in passing that the latter property is inherent in most, if not all,
mechanisms of the generation of cosmological perturbations.

Equation~\eqref{jul22-5} is exactly the same as equation for
minimally coupled massless scalar field in the
de~Sitter background. Nevertheless, let us briefly discuss its solutions.
At early times, when the inequality \eqref{jul22-3} is satisfied,
$\delta \chi_2$ is free massless quantum field,
\[
\delta \chi_2({\bf x}, \eta) = \int~\frac{d^3k}{(2\pi)^{3/2}
\sqrt{2k}}~\left( \delta \chi_2^{(-)}({\bf k},
{\bf x}, \eta)  \hat{A}_{\bf k} + h.c.\right)\; ,
\]
whose modes are
(we keep the dependence on ${\bf x}$ for future convenience)
\be
\delta \chi_2^{(-)} ({\bf k},
{\bf x}, \eta)=
%\frac{1}{(2\pi)^{3/2}
%\sqrt{2k}}\cdot
\mbox{e}^{i {\bf k x} - ik\eta}
%\cdot
%\mbox{e}^{-ik \eta}
\; .
\label{jul22-4}
%\nonumber
\ee
Here  $\hat{A}_{\bf k}$ and  $\hat{A}_{\bf k}^\dagger$  are aniihilation
and creation operators obeying the standard commutational relation,
$[\hat{A}_{\bf k},\hat{A}_{\bf k^\prime}^\dagger] = \delta({\bf k} - {\bf
k^\prime})$. It is  natural to assume that the field $\delta \chi_2$ is
initially in its vacuum state.

The rolling background $\chi_c (\eta)$ produces an effective ``horizon''
for the perturbations $\delta \chi_2$. The oscillations \eqref{jul22-4}
terminate when the mode exits the ``horizon'', i.e., at $k(\eta_* - \eta)
\sim 1$. The solution to eq.~\eqref{jul22-5}
%that oscillates at early times according to
with the initial condition
 \eqref{jul22-4} is
\be
\delta \chi_2^{(-)} ({\bf k},
{\bf x}, \eta)= \mbox{e}^{i {\bf kx} - ik\eta_*}\cdot F(k, \eta_* - \eta)
\; ,
\label{jul25-5}
\ee
where
\be
F(k, \xi) = - \sqrt{\frac{\pi}{2}k \xi}~
H^{(1)}_{3/2} (k\xi)
\label{jul25-6}
\ee
and $H^{(1)}_{3/2}$ is the Hankel function. In the late-time
super-''horizon'' regime, when $k(\eta_* - \eta) \ll 1$, one has
\be
F(k, \eta_* - \eta) = \frac{i}{k(\eta_* - \eta)} \; .
\label{jul27-40}
\ee
Hence, the super-''horizon'' perturbations of the phase
$\delta \theta \equiv \delta \chi_2/\chi_c$ are time-independent,
%\marginpar{\bf \tiny $h$ aded in \eqref{jul25-33}}
\be
\delta \theta ({\bf x}) =\frac{\delta \chi_2({\bf x}, \eta)}{\chi_c
(\eta)} = ih \int~\frac{d^3k}{4\pi^{3/2}k^{3/2}}~
%\left(
 \mbox{e}^{i{\bf kx} - i k\eta_*}
\hat{A}_{\bf k}
%\right)
+ h.c.
\label{jul25-33}
\ee
This expression describes Gaussian random field (cf. ref.~\cite{PolStar})
whose power spectrum is given by \eqref{jul21-1}.

\subsection{Converting perturbations of
phase into adiabatic perturbations}
\label{Reprocess}

For the sake of completenetss, let us briefly discuss two possible
ways to reprocess the perturbations of the phase $\delta \theta$
into the adiabatic perturbations.

\begin{figure}[tb!]
\begin{center}
\includegraphics[width=0.7\textwidth,angle=0]{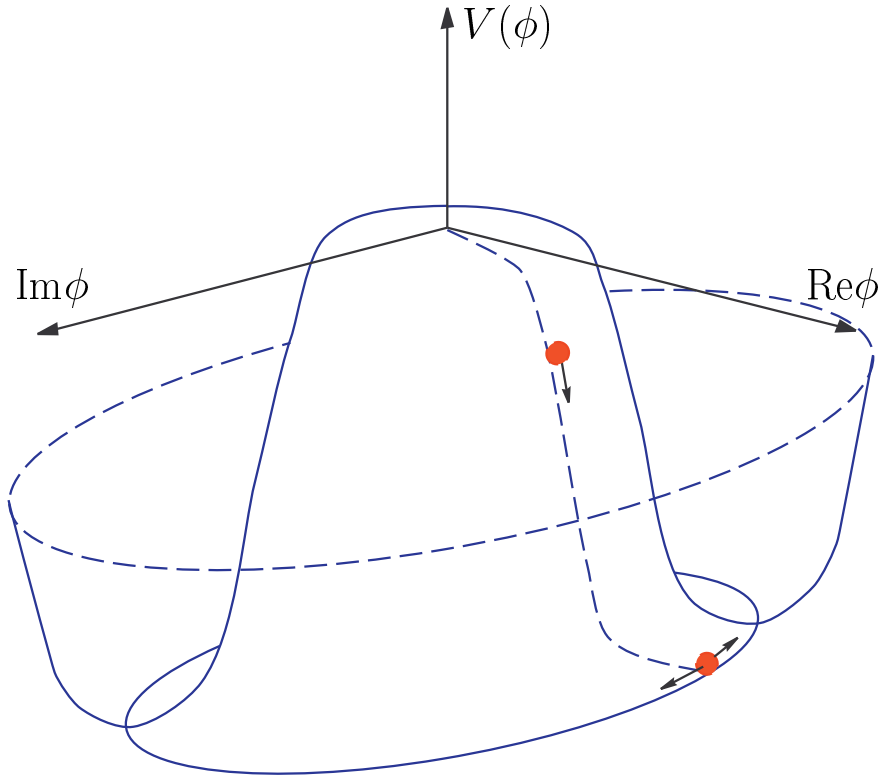}
\end{center}
\caption{The scalar potential in the pseudo-Nambu-Goldstone
scenario.
\label{fig2}}
\end{figure}

One way is to make use of the curvaton
mechanism~\cite{Linde:1996gt} specified to pseudo-Nambu-Goldstone
curvaton~\cite{Dimopoulos:2003az}. Namely,
let us assume that $\theta$ is actually pseudo-Nambu-Goldstone
field. Generically, conformal rolling ends up at a slope of its potential,
see Fig.~\ref{fig2}.
The field $\theta$ together with its perturbations remains
 frozen out until the time at the hot Big Bang epoch
when the Hubble parameter becomes of the order of the mass parameter of
this field. Then $\theta$ starts to oscillate about the minimum of
its potential. Provided this field interacts with conventional particles,
its oscillations eventually decay, and their energy density is
transfered to the cosmic plasma. At that epoch, the perturbations of
the energy density of the field $\theta$ are converted into
the adiabatic perturbations. The shape of the
resulting adiabatic power spectrum is the same as that of the initial
spectrum of $\delta \theta$, i.e., it is flat in the case of exact conformal
invariance at the rolling stage. Generically,
the amplitude of the adiabatic perturbations
is of order
\[
\sqrt{{\cal P}_\zeta} \sim r \frac{\sqrt{{\cal P}_{\delta \theta}}}{\theta_c}
= r \frac{h}{2\pi \theta_c}\; ,
\]
where $r$ is the ratio of the energy density of the field $\theta$ to
the total energy densiy at the time the $\theta$-oscillations decay,
and $\theta_c$ is the distance from the landing point of rolling to the
minimum of the potential.

The pseudo-Nambu-Goldstone mechanism produces the
non-Gaussianity of local form
in the
adiabatic perturbations. For generic values of the phase at landing,
$\theta_c \sim \pi/2$, non-observation of the
non-Gaussianity~\cite{Komatsu:2010fb} implies
\[
   r \gtrsim 10^{-2} \; ,
\]
so that the correct scalar amplitude is obtained for
\be
  h \lesssim 10^{-2} \; .
\label{jul27-10}
\ee
Thus, the scenario with the pseudo-Nambu-Goldstone mechanism
of conversion of the phase perturbations into the adiabatic ones
is viable for quite small scalar self-coupling only.

Another possibility is the modulated decay of heavy
particles~\cite{Dvali:2003em,Dvali:2003ar,Vernizzi:2003vs}.
One assumes that
the phase field $\theta$ interacts with some heavy particles
in such a way that the masses and/or widths of the latter
depend on $\theta$,
\be
M= M_0 + \epsilon_M \theta\;\;\;\;  \mbox{and/or}\;\;\;
\Gamma = \Gamma_0 + \epsilon_\Gamma \theta \; .
\label{jul23-1}
\ee
One assumes further that these particles survive at the hot Big Bang
epoch until they
are non-relativistic and dominate
the cosmological expansion.
When these particles decay, the perturbations
in $\theta$, and hence in
$M$ and/or $\Gamma$,
induce adiabatic perturbations,
\[
\zeta \sim \frac{\delta M}{M}
= \frac{\epsilon_M \delta \theta}{M_0 + \epsilon_M \theta_c}
\;\;\;\;  \mbox{and/or} \;\;\; \zeta \sim \frac{\delta \Gamma}{\Gamma}
= \frac{\epsilon_\Gamma \delta \theta}{\Gamma_0 + \epsilon_\Gamma
\theta_c} \; .
\]
The shape of the
adiabatic power spectrum is again the same as that of the initial
power spectrum of $\delta \theta$.

The modulated decay mechanism
also induces non-Gaussianity in the adiabatic perturbations.
However, once the dependence of the mass/width on $\theta$ is linear,
as written in \eqref{jul23-1},
the induced non-Gaussian part of the adiabatic
perturbations is of order
\[
\l \delta M/M \r^2 ,
\;
%\;
\l \delta \Gamma/\Gamma \r^2 \sim \zeta^2 \; .
\]
In other words, irrespectively of the value of
the coupling $h$, the non-Gaussianity parameter is fairly small
(see refs.~\cite{Dvali:2003ar,Vernizzi:2003vs} for details),
\[
f_{NL} \sim 1 \; ,
\]
in comfortable agreement with the existing limit~\cite{Komatsu:2010fb}.
Thus, the modulated decay mechanism by itself does not imply
any bound on $h$.

\subsection{Perturbations of modulus}
\label{modulusperturbations}

Let us now come back to the conformal rolling stage and consider
the radial perturbations --- perturbations
of the modulus of the field $\chi$, or, with
our convention of real backgronud $\chi_c$, perturbations of the real part
\[
  \chi_1 \equiv \frac{\mbox{Re}~\chi}{\sqrt{2}} \; .
\]
At the linearized level, they obey the following equation,
%in momentum representation,
\be
(\delta \chi_1)^{\prime \prime}
- \d_i\d_i~ \delta \chi_1 - 6 \chi_c^2 \delta \chi_1 \equiv
(\delta \chi_1)^{\prime \prime}
- \d_i\d_i~ \delta \chi_1 - \frac{6}{(\eta_* - \eta)^2} \delta \chi_1  = 0
\; . \nonumber \ee Its solution that tends to properly normalized mode of
free quantum field as $k(\eta_* -\eta) \to \infty$ is
\[
\delta \chi_1 = \mbox{e}^{i{\bf kx}-ik\eta_*} \cdot
\frac{-i}{4\pi} \sqrt{\frac{\eta_* - \eta}{2}}
H_{5/2}^{(1)} \left[k (\eta_* - \eta) \right] \cdot \hat{B}_{\bf k} +
h.c.\; ,
\]
where $\hat{B}_{\bf k}$, $\hat{B}_{\bf k}^\dagger$
is another set of annihilation and
creation operators. At late times, when $k(\eta_* - \eta) \ll 1$
(super-''horizon'' regime), one has
\[
\delta \chi_1 = \mbox{e}^{i{\bf kx}-ik\eta_*} \cdot
\frac{3}{4\pi^{3/2}} \frac{-1}{k^{5/2} (\eta_* - \eta)^2} \cdot
\hat{B}_{\bf k} + h.c.\; .
\]
Hence, the super-''horizon''
perturbations of the modulus have red power spectrum \eqref{jul23-2}.

The dependnce $\delta \chi_1 \propto (\eta_* - \eta)^{-2}$ is naturally
interpreted in terms of the local shift of the ``end time'' parameter
$\eta_*$.
Indeed, with the background field given by \eqref{jul22-2},
the sum $\chi_c + \delta \chi_1$,
%giving
i.e.,
the
radial field including perturbations, is the linearized
form of
\be
\chi_c [\eta_* ({\bf x}) - \eta] = \frac{1}{h[\eta_* ({\bf x}) - \eta]} \;
,
\label{jul24-1}
\ee
where
\be
\eta_* ({\bf x}) = \eta_* + \delta \eta_* ({\bf x})
\label{jul25-1}
\ee
and
\be
\delta \eta_* ({\bf x}) = \frac{-3h}{4\pi^{3/2}}\int~\frac{d^3k}{k^{5/2}}
\l \mbox{e}^{i{\bf kx} - ik\eta_*} \cdot
\hat{B}_{\bf k}
+ h.c. \r \; .
\label{jul27-1}
\ee
So, the infrared radial modes  modify the effective background
by
transforming the ``end time'' parameter $\eta_*$  into random field
that slowly varies in space,\footnote{There are corrections to
eq.~\eqref{jul24-1} of order $\d_i \d_j \eta_* ({\bf x})$ and
$[\d_i \eta_* ({\bf x})]^2$, see sections~\ref{h}
and \ref{Newsec}.} as given in eqs.~\eqref{jul24-1},
\eqref{jul25-1}. Clearly,
this observation is valid beyond the linear approximation:
once the spatial scale of variation of $\chi_1 ({\bf x},\eta)$
exceeds the ``horizon'' size, spatial gradients in
eq.~\eqref{jul24-2} are negligible, and
the late-time
solutions to the full non-linear field equation have locally
one and the same form \eqref{jul22-2}, modulo slow variation of
$\eta_*$ in space.

%\marginpar{\bf \tiny Paragraph added}
It is worth noting that the infrared modes contribute both to the
field $\delta \eta_* ({\bf x})$ itself and to its spatial derivative.
The contribution of the modes which are superhorizon today,
i.e., have momenta $k\lesssim H_0$, to
the variance of the latter is given by
\be
\langle \d_i \eta_* ({\bf x})
\d_j \eta_* ({\bf x}) \rangle_{k\lesssim H_0}
= \delta_{ij} \cdot
\frac{3h^2}{4\pi} \int_{k\lesssim H_0} \frac{dk}{k}
= \delta_{ij} \cdot
\frac{3h^2}{4\pi} \log \frac{H_0}{\Lambda} \; ,
\label{oct5-3}
\ee
where $\Lambda$ is the infrared cutoff
which parametrizes our ignorance of the dynamics at the
beginning of the conformal rolling stage.

\section{Effect of infrared radial modes
on perturbations of  phase: first order in $h$}
\label{twoorders}

%\marginpar{\bf \tiny Title of Sec. changed}
The main purpose of this paper is to understand how the interaction
with the infrared radial modes  affects the properties of the
perturbations of the phase $\delta \theta$. To this end, we consider
perturbations of the imaginary part $\delta \chi_2$, whose wavelengths
are much smaller than the scale of the spatial variation of the modulus.
Because of the separation of scales, perturbations $\delta \chi_2$
can still be treated in the linear approximation, but now in
the background \eqref{jul24-1}.

%Instead of eq.~\eqref{jul22-5}, we now have
%\be
%(\delta \chi_2)^{\prime \prime}
%+ k^2 \delta \chi_2
%- \d_i \d_i \; \delta \chi_2
%- \frac{2}{[\eta_*({\bf x}) - \eta]^2} \delta \chi_2  = 0 \; .
%\label{jul25-3}
%\nonumber
%\ee
Since our concern is the infrared part of $\eta_*({\bf x})$, we
make use of the spatial gradient expansion, consider, for the time being,
a region near the
origin and write
\be
\eta_* ({\bf x}) = \eta_* (0) - v_i x_i + \dots \; ,
\label{jul24-3}
\ee
where
\[
v_i = -\d_i \eta_*({\bf x})\vert_{{\bf x}=0} \; ,
\]
and dots denote higher order terms in ${\bf x}$.
Importantly,  the field $\d_i \d_j \eta_* ({\bf x})$ has blue
power spectrum, unlike $\eta_* ({\bf x})$ and $\d_i \eta_* ({\bf x})$,
so the major effect of the infrared modes is accounted for
by considering the two terms of the gradient expansion written
explicitly
in \eqref{jul24-3}. In this section we work at this, first order
of the gradient expansion. Furthermore,
%\marginpar{\bf \tiny Changed to end of paragraph}
we assume in what follows that
\be
|{\bf v}| \ll 1 \; ,
\label{jul25-2}
\ee
and in this section we
neglect corrections of order ${\bf v}^2$.
 The expansion in $|{\bf v}|$ is legitimate, since
the field ${\bf v} ({\bf x})$ has flat power spectrum, so
the fluctuation of ${\bf v}$ is of order $h^2 |\log \Lambda|$, where
$\Lambda$ is the infrared cutoff, and it is small for small $h$ and not
too large  $|\log \Lambda|$. In other words, the expansion in
$|{\bf v}|$ is the expansion in $h$, modulo infrared logarithms.
We postpone to section~\ref{Newsec} the analysis of the leading effect
that occurs at the order ${\bf v}^2$.

%The assumption \eqref{jul25-2} can be relaxed
%by making use of fully relativistic formulas, see footnotes below.

Keeping the two terms in \eqref{jul24-3} only, we have,
instead of eq.~\eqref{jul22-5},
\be
(\delta \chi_2)^{\prime \prime}
%+ k^2 \delta \chi_2
- \d_i \d_i \; \delta \chi_2
- \frac{2}{[\eta_*(0) - {\bf vx} - \eta]^2} \delta \chi_2  = 0 \; .
\label{jul25-3}
%\nonumber
\ee
We
observe that
the denominator in the expression for the background field
\be
\chi_c = \frac{1}{h[\eta_* (0) - \eta - {\bf vx}]}
\label{jul25-20}
\ee
contains the combination
$\eta_* (0) - (\eta + {\bf vx})$.
We interpret this as the local time shift and Lorentz boost
of the original background \eqref{jul22-2}: the effective background
is homogeneous in a reference frame (in conformal coordinates)
other than the cosmic frame where the metric is spatially
homogeneous. Note that the field
\eqref{jul25-20} is a solution to the field equation
%\marginpar{\bf \tiny Changed}
%\footnote{If
%terms of order ${\bf v}^2$ and higher are not
%neglected, the function \eqref{jul25-20}
%is no longer
%a solution to the field equation \eqref{jul24-2}. Instead, the solution
%is (still keeping two terms only in the gradient expansion)
%\be
%\chi_c = h^{-1} \left(\frac{\eta_* (0) - \eta - {\bf vx}}{\sqrt{1-v^2}}
%\right)^{-1} \; .
%\label{jul25-10}
%\ee
%So, for arbitrary $|{\bf v}|$ it is appropriate to study the solutions
%to eq.~\eqref{jul25-11} in this background.}
\eqref{jul24-2} in our
approximation.
Our interpretation suggests that the solutions to eq.~\eqref{jul25-3}
can be obtained by time translation and Lorentz boost of the original
solution \eqref{jul25-5}, \eqref{jul25-6}. Indeed, it is straightforward
to see
%\marginpar{\bf \tiny Changed}
that to the first order in
${\bf v}$, the solution to eq.~\eqref{jul25-3} obeying the initial
condition \eqref{jul22-4} is \be \delta \chi_2^{(-)} ({\bf k}, {\bf x},
\eta)= \mbox{e}^{i {\bf q}({\bf x} + {\bf v}\eta) - iq\eta_* (0)} \cdot
F(q, \eta_*(0) - \eta - {\bf vx}) \; ,
\label{jul25-7}
\ee
where the function $F$ is still defined by \eqref{jul25-6}, the
Lorentz-boosted momentum is
\be
 {\bf q} = {\bf k} + k{\bf v} \; , \;\;\;\;
 q = |{\bf q}| = k + {\bf kv} \; ,
\label{jul25-31}
\ee
and it is understood that terms of order ${\bf v}^2$ must be
neglected.
%\footnote{For arbitrary $|{\bf v}|$,
% the appropriate background is \eqref{jul25-10}.
%The solution to eq.~\eqref{jul25-11} in this background, that
%obeys the initial condition \eqref{jul22-4}, is
%\be
%\delta \chi_2^{(-)} ({\bf k},
%{\bf x}, \eta)= \mbox{exp}\left(
%i q_{||}\frac{x_{||} + v\eta}{\sqrt{1-v^2}}  + i {\bf q}^T {\bf x}^T
%- iq\frac{\eta_* (0)}{\sqrt{1-v^2}}\right)
%\cdot F\left(q, \frac{\eta_*(0) - \eta - {\bf vx}}{\sqrt{1-v^2}}\right)
%\; , \label{jul25-12}
%\ee
%where the Lorentz-boosted momentum is, as usual,
%$q_{||} = \frac{k_{||} + kv}{\sqrt{1-v^2}}$, ${\bf q}^T = {\bf k}^T$,
%$q= \frac{k + k_{||}v}{\sqrt{1-v^2}}$, and notations $||$ and $T$
%refer to components, parallel and normal to ${\bf v}$, respectively.
%The rest of the analysis in this section can be generalized to the case
%of arbitrary $|{\bf v}|$ by making use of the formulas \eqref{jul25-10}
%and \eqref{jul25-12}; the result is the same.}.
The solution
\eqref{jul25-7} can be cast into the following form,
\be
\delta \chi_2^{(-)} ({\bf k},
{\bf x}, \eta)= \mbox{e}^{i {\bf k}{\bf x}  - ik\eta_* ({\bf x})
- i{\bf kv}[\eta_*({\bf x}) - \eta]}
\cdot F(q, \eta_*({\bf x}) - \eta) \; ,
\label{jul25-30}
\ee
where
\[
v_i = - \d_i \eta_* ({\bf x}) \; .
\]
This form, still valid to the first order in the gradient expansion
and to the linear order in ${\bf v}$, does not make any reference to
the origin of the coordinate frame
and can be used for arbitrary ${\bf x}$. We consider corrections
of order $\d_i \d_j \eta_* ({\bf x})$ and ${\bf v^2}$
%\marginpar{\bf \tiny Changed}
to this solution in
sections~\ref{h} and \ref{Newsec}, respectively.

We find from eqs.~\eqref{jul25-20} and
\eqref{jul25-7} (or, equivalently, \eqref{jul24-1} and
\eqref{jul25-30}) that the perturbations of the phase
again freeze out as $k[\eta_*({\bf x}) - \eta] \to 0$, now at
\be
\delta \theta ({\bf x}) =\frac{\delta \chi_2({\bf x})}{\chi_c (\bf x)}
= i \int~\frac{d^3k}{\sqrt{k}}\frac{h}{4\pi^{3/2} q}~
%\left(
\mbox{e}^{i{\bf kx} - i k\eta_*({\bf x})}
\hat{A}_{\bf k}
%\right)
+ h.c.\; ,
\label{jul25-32}
\ee
where the relation between ${\bf k}$ and ${\bf q}$ is still
given by eq.~\eqref{jul25-31}. We recall that $\eta_* ({\bf x})$
here is a realization of infrared random field.
The formula \eqref{jul25-32} implies that to the
first order of the gradient expansion we limit ourselves in this section,
the properties of
the random field $\delta \theta$ are the same as those of the linear
field \eqref{jul25-33}. Consider, e.g., the two-point correlation
function
\be
\langle \delta \theta ({\bf x}_1) \; \delta \theta ({\bf x}_2)
\rangle = \int~\frac{d^3k}{k}\frac{h^2}{16\pi^{3} q^2}
\mbox{e}^{i{\bf k}({\bf x}_1-{\bf x}_2)   - i k[\eta_*({\bf x}_1)
- \eta_* ({\bf x}_2)]} + c.c.
\label{jul25-35}
\ee
Infrared modes of $\eta_*({\bf x})$ we consider have momenta
much lower than $k$, i.e., the spatial scale of their
variation much exceeds $|{\bf x_1}-{\bf x}_2|$. Hence,
to the first order in the gradient expansion we have
$\eta_*({\bf x}_1)
- \eta_* ({\bf x}_2)=-{\bf v}({\bf x_1} - {\bf x}_2)$,
and ${\bf v}$ is independent of ${\bf x}$. Therefore, the two-point
function is
\[
\langle \delta \theta ({\bf x}_1) \; \delta \theta ({\bf x}_2)
\rangle = \int~\frac{d^3k}{k}\frac{h^2}{16\pi^{3} q^2}
\mbox{e}^{i{\bf q}({\bf x}_1-{\bf x}_2)} + c.c.
\]
We now change the integration varible from ${\bf k}$ to ${\bf q}$,
recall that the integration measure $d^3k/k$ is Lorentz-invariant,
and obtain
\[
\langle \delta \theta ({\bf x}_1) \; \delta \theta ({\bf x}_2)
\rangle = h^2 \int~\frac{d^3q}{16\pi^{3} q^3}
\mbox{e}^{i{\bf q}({\bf x}_1-{\bf x}_2)} + c.c.
\]
This is precisely the two-point correlation function of the
linear field \eqref{jul25-33}.

The latter argument is straightforwardly generalized to
multiple correlators: for a given realization of the random
field $\eta_* ({\bf x})$, they are all expressed in terms
of the two-point correlation function
\eqref{jul25-35}. In other words, the infrared effects
are removed by the field redefinition,
\be
\hat{\mathcal A}_{\bf q}
 = \mbox{e}^{-ik\eta_* (0)}\sqrt{\frac{k}{q}} \hat{A}_{\bf k} \; ,
\label{oct5-4}
\ee
where ${\bf k}$ and ${\bf q}$ are still related by \eqref{jul25-31}.
The operators $\hat{\mathcal A}_{\bf q}$,
$\hat{\mathcal A}_{\bf q}^\dagger$ obey the standard
commutational relations, while in our approximation,
the field \eqref{jul25-32}, written in terms
of these operators, coincides with the linear field \eqref{jul25-33}.
We conclude that the
%\marginpar{\bf \tiny Changed}
infrared radial modes are, in fact, not particularly
dangerous, as they do not
grossly affect the properties of the field $\delta \theta$.

\section{Statistical anisotropy}
\label{anisotropy}

%\marginpar{\bf \tiny Changed to end of paragraph}

\subsection{First order in $h$}
\label{h}

Let us continue with the analysis at the first order in $h$.
In this approximation,
the non-trivial effect of the large wavelength perturbations
$\delta \eta_* ({\bf x})$
on the perturbations of the phase, and hence on the resulting
adiabatic perturbations, occurs for the first time at the second order in the
gradient expansion, i.e., at the order $\d_i \d_j \eta_*$.
Let us concentrate on the effect of the modes of
$\delta \eta_*$ whose present wavelengths exceed the present Hubble
size. We are dealing with one realization of the random field
$\delta \eta_*$, hence at the second order of the gradient
expansion, $\d_i \d_j \eta_*$ is merely
a tensor, constant throughout the visible Universe.
In this section we calculate the statistical anisotropy associated with
this tensor.

To this end, we make use of the perturbation theory in
$\d_i \d_j \eta_*$. In the first place, we have to find the
background, since
the function
\eqref{jul24-1} is no longer the solution to the field
equation~\eqref{jul24-2} at the second order in the gradient expansion. We write
\be
\chi_c = \chi_c^{(1)} + \chi_c^{(2)} \; ,
\label{jul26-1}
\ee
where
\be
\chi_c^{(1)}  = \frac{1}{h[\eta_* ({\bf x}) - \eta]} \; ,
\label{jul26-3}
%\nonumber
\ee
and $\chi_c^{(2)} = O(\d_i \d_j \eta_*)$. We plug the expression
\eqref{jul26-1} into eq.~\eqref{jul24-2}, linearize in
$\d_i \d_j \eta_*$ and obtain the following equation for the correction,
\be
\chi_c^{(2)\, \prime \prime} - \frac{6}{(\eta_* - \eta)^2} \chi_c^{(2)}
= - \frac{\d_i \d_i \eta_*}{h(\eta_* - \eta)^2} \; ,
\label{jul26-2}
\ee
where we again neglect terms of order ${\bf v}^2$.
Clearly, the correction to the background depends only on
the scalar $\d_i \d_i \eta_*$, so it does not yield statistical
anisotropy. Nevertheless, we keep this correction in what follows.
%to cross check
%the consistency of the entire approach.

The  general solution to eq.~\eqref{jul26-2} is
\[
\chi_c^{(2)} = C_1 (\eta_* - \eta)^3 + \frac{C_2}{(\eta_* - \eta)^2} +
\frac{1}{6h} \d_i \d_i \eta_* \; .
\]
The first term in the right hand side
is irrelevant in the super-''horizon'' regime,
the second term is merely a shift of $\eta_*$ in the leading
order expression \eqref{jul26-3}, so the non-trivial correction is
given by the third term. The combination entering
eq.~\eqref{jul25-11} for the perturbations of the imaginary part
is now given by (we consistently work at the linear order
in $\d_i \d_j \eta_*$)
\be
2h^2 \chi_c^2 = \frac{2}{(\eta_* ({\bf x}) - \eta)^2} + \frac{2}{3}
\frac{\d_i \d_i \eta_*}{\eta_* - \eta} \; .
\label{jul26-13}
\ee
Note that we can set $\eta_* = \mbox{const}$ in the second term,
since we
%\marginpar{\bf \tiny Changed}
neglect corrections of order $\d_i \d_j \eta_*
\cdot \d_k \eta_*$.

Let us now obtain the solution to eq.~\eqref{jul25-11}
to the first order in $\d_i \d_j \eta_*$. The initial condition is
still given by eq.~\eqref{jul22-4}. We search for the solution
in the following form,
\be
\delta \chi_2^{(-)} ({\bf k},
{\bf x}, \eta)= \mbox{e}^{i {\bf k}{\bf x}  - ik\eta_* ({\bf x})
- i{\bf kv}[\eta_*({\bf x}) - \eta]}
\cdot \left[F(q, \eta_*({\bf x}) - \eta) + F^{(2)}(q, \eta_*({\bf x}) -
\eta) \right]\; ,
%\label{jul25-30}
\nonumber
\ee
where the leading term $F$ is again defined by \eqref{jul25-6} and
$F^{(2)}$ is proportional to $\d_i \d_j \eta_*$.
We expand eq.~\eqref{jul25-11} in  $\d_i \d_j \eta_*$ and obtain to
the linear order
\be
F^{(2)\, \prime \prime} + k^2 F^{(2)} - \frac{2}{\xi^2} F^{(2)}
= \d_i \d_i \eta_* \cdot S + k_i k_j \; \d_i \d_j \eta_* \cdot T \; ,
\label{jul26-11}
\ee
where $\xi = \eta_* - \eta$,
\begin{align}
S &= - ikF + \frac{\d F(k,\xi)}{\d \xi} + \frac{2}{3\xi} F \; ,
\label{jul26-12}
\\
T &= -2 F \xi - 2i \frac{\d F(k, \xi)}{\d k} =
- \frac{2}{k^2 \xi} \mbox{e}^{ik\xi} \; .
\label{jul26-10}
\end{align}
After calculating the right hand side of eq.~\eqref{jul26-11}
we set ${\bf q} = {\bf k}$ and ${\bf v}=0$, since we neglect
the terms of order
%\marginpar{\bf \tiny Changed}
$\d_i \d_j \eta_* \cdot \d_k \eta_*$. Note that
the last term in \eqref{jul26-12} comes from the correction to
the background, see  eq.~\eqref{jul26-13}, and that the
last expression in \eqref{jul26-10} is obtained by using the
explicit form of the Hankel function $H_{3/2}^{(1)}$.

The soultion $F^{(2)}$ should vanish as $\eta_* - \eta \to \infty$.
Hence, it is given in terms of the retarded Green's function
(recall that $\xi^\prime > \xi$ corresponds to
$\eta^\prime < \eta$)
\be
G(\xi, \xi^\prime) =
\frac{\pi \sqrt{\xi \xi^\prime}}{2} \cdot
\Theta(\xi^\prime - \xi) \cdot \left[ J_{3/2}(k\xi) N_{3/2} (k \xi^\prime)
- N_{3/2} (k\xi) J_{3/2}(k \xi^\prime)\right] \; ,
\label{jul26-20}
\ee
where $J_{3/2}$ and $N_{3/2}$ are the Bessel functions. Namely,
\be
F^{(2)} (\xi) =
\int_\xi^\infty~d\xi^\prime~G(\xi, \xi^\prime)
\left[ \d_i \d_i \eta_* \cdot S (\xi^\prime) + k_i k_j \; \d_i \d_j \eta_* \cdot T
(\xi^\prime) \right] \; .
\label{jul26-21}
\ee
We are interested in the behavior of this solution in the super-''horizon''
regime, $k\xi \to 0$. Since the most singular behavior of
$S$ and $T$ at small $\xi$ is $\xi^{-2}$, the first term in
square brackets in \eqref{jul26-20} gives finite contribution to
the integral \eqref{jul26-21} as $\xi \to 0$, while the second term yields
$F^{(2)} (\xi) \propto \xi^{-1}$. Hence, the behavior of the correction
$F^{(2)} (\xi)$ at small $\xi$ is the same as that of the leading term $F(\xi)$,
so the phase perturbations $\delta \theta$, the correction included,
are constant in time in the super-''horizon'' regime. This is a
consistency check of our approach.

Since we treat $\d_i \d_j \eta_*$ as constant over the visible
Universe, the power spectrum of the phase perturbations
contains a part linear\footnote{Had we considered the perturbations
$\delta \eta_*$ whose wavelengths are much shorter than the present
Hubble size, it would be appropriate to perform ensemble
averaging.
In that case the linear part would average out.}
 in $\d_i \d_j \eta_*$. This linear part
 comes from the interference of $F^{(2)}$ and $F$.
In the super-''horizon'' regime, $F$ is pure imaginary, see
\eqref{jul27-40}. Hence, we are interested in the behavior
of the imaginary part of $F^{(2)}$ as $\xi \to 0$.
This behavior is straightforward to obtain.
Namely, for $T$-term in \eqref{jul26-21} we write
\begin{align}
\mbox{Im}~F^{(2)\, T} \vert_{\xi \to 0}
&=  - \frac{\pi \sqrt{\xi}}{2} N_{3/2} (k\xi)
\cdot \int_0^\infty ~d\xi^\prime
\sqrt{\xi^\prime} J_{3/2}(k\xi^\prime)
\left( - \frac{2\sin k\xi^\prime}{k^2\xi^\prime} \right)
\cdot k_ik_j\, \d_i \d_j \eta_*
\nonumber\\
&= - \frac{\pi}{2}
\frac{1}{k^2\xi} \cdot \frac{k_ik_j}{k^2} \d_i \d_j \eta_* \; .
\nonumber
\end{align}
Performing similar calculation for $S$-term in \eqref{jul26-21},
we obtain that in the super-''horizon'' regime
\[
F + F^{(2)} = \frac{i}{q(\eta_* - \eta)}
\l 1 - \frac{\pi}{2k}  \cdot \frac{k_ik_j}{k^2} \d_i \d_j \eta_*
+ \frac{\pi}{6k} \d_i \d_i \eta_* \r \; .
\]
The two non-trivial terms in parenthesis give the correction to the power
spectrum of the phase perturbations due to the radial modes whose
whavelengths exceed the present Hubble size. The same correction
is characteristic of the adiabatic perturbations, so we have finally
\be
{\cal P}_\zeta = A_\zeta \left[ 1 - \frac{\pi}{k} \cdot
\frac{k_ik_j}{k^2} \l \d_i \d_j \eta_* - \frac{1}{3} \delta_{ij}
\d_k \d_k \eta_* \r \right] \; ,
\label{oct5-1}
\ee
where the adiabatic amplitude $A_\zeta$ is independent of $k$
within our aproximations.
Notably, the angular average of the correction vanishes, so
we are dealing
with the statistical anisotropy proper.

Neither the magnitude nor the exact form of the tensor
$\d_i \d_j \eta_* - (1/3) \delta_{ij}
\d_k \d_k \eta_*$ can be unambiguously predicted
because of the cosmic variance.
To estimate the strength of the statistical anisotropy, let us
consider the variance
\[
\langle (\d\d\eta_*)^2 \rangle \equiv
\langle \l \d_i \d_j \eta_* - \frac{1}{3} \delta_{ij}
\d_k \d_k \eta_*\r \cdot \l\d_i \d_j \eta_* -  \frac{1}{3} \delta_{ij}
\d_k \d_k \eta_* \r\rangle_{k \lesssim H_0} \; ,
\]
where the notation reflects
the fact that we take into account only those modes whose
present wavelengths exceed the present Hubble size.
We make use of \eqref{jul27-1} and obtain
\be
\langle (\d\d\eta_*)^2 \rangle = \frac{9h^2}{16\pi^2}
\int_{k\lesssim H_0}~\frac{dk}{k^5}
\cdot \frac{2k^4}{3} \simeq \frac{3h^2}{4\pi^2} H_0^2 \; .
\label{oct5-2}
\ee
In this way
%\marginpar{\bf \tiny Changed}
we arrive at the first non-trivial term in \eqref{jul22-6}.
Higher orders in the gradient expansion give contributions to the
statistical anisotropy which are suppressed by extra factors of
$H_0/k$.

\subsection{Order $h^2$: contribution of deep infrared modes}
\label{Newsec}

%\marginpar{\bf \tiny subsection added}
Let us now turn to the statistical anisotropy at the second order in
$h$. The major contribution at this order is proportional to
$\d_i \eta_* \d_j \eta_* \equiv v_i v_j$. Indeed, the overall time shift
$\eta_*(0)$ is irrelevant, while the terms
involivng higher derivative combinations like $\d_i\d_j \eta_* \d_i \eta_*$
are suppressed by powers of $H_0/k$, cf.~\eqref{oct5-1}, \eqref{oct5-2}.
Furthermore, the ``velocity'' ${\bf v}$ is enhanced, albeit only
logarithmically, by the deep infrared effects, see \eqref{oct5-3}.
Hence, to extract the major contribution to the statistical anisotropy
at the order $h^2$, we use the two terms of the derivative expansion,
explicitly written in \eqref{jul24-3}.

Once the terms of order ${\bf v}^2$ are not
neglected, the function \eqref{jul25-20}
is no longer
a solution to the field equation \eqref{jul24-2}. Instead, the solution
is (keeping two terms only in the gradient expansion)
\be
\chi_c = \frac{1}{h \gamma \left[\eta_* (0) - \eta - {\bf vx}
\right]} \; .
\label{jul25-10}
\ee
So,  it is appropriate to study the solutions
to eq.~\eqref{jul25-11} in this background. It is straightforward to see
that
the solution  that
obeys the initial condition \eqref{jul22-4} is
\be
\delta \chi_2^{(-)} ({\bf k},
{\bf x}, \eta)= \mbox{e}^{
i q_{||}\gamma(x_{||} + v\eta)  + i {\bf q}^T {\bf x}^T
- iq\gamma \eta_* (0)}
\cdot F\left[q, \gamma(\eta_*(0) - \eta - {\bf vx})\right] \; ,
\label{jul25-12}
\ee
where the Lorentz-boosted momentum is, as usual,
$q_{||} = \gamma(k_{||} + kv)$, ${\bf q}^T = {\bf k}^T$,
$q= \gamma(k + k_{||}v)$, $\gamma =(1-v^{2})^{-1/2}$, and notations $||$
and $T$ refer to components which are parallel and normal to ${\bf v}$,
respectively. The interpretation of this solution is again that it is the
Lorentz-boosted solution \eqref{jul25-5}.

According to the scenario discussed in this paper,
the phase perturbations $\delta \theta$ freeze out at the
hypersurface
\[
\eta = \eta_* (0) - {\bf vx} \equiv \eta_*(0) - v x_{||}
\]
and then stay constant in the cosmic time $\eta$. It follows from
\eqref{jul25-10} and \eqref{jul25-12} that at this hypersurface
and later, the perturbations of the phase are
\be
\delta \theta ({\bf x})
=  \int~\frac{d^3q}{\sqrt{q}}\frac{h}{4\pi^{3/2} q}~
\mbox{e}^{
i \gamma^{-1}q_{||}x_{||}   + i {\bf q}^T {\bf x}^T}
\hat{{\mathcal A}}_{\bf q}
%\right)
+ h.c.\; ,
\label{oct5-5}
\ee
where the operators $\hat{{\mathcal A}}_{\bf q}$ are defined by
\eqref{oct5-4} and obey the standard commutational relations,
and we have omitted irrelevant constant phase factor.
We see that to the order ${\bf v}^2$, the effect of the deep
infrared modes does not disappear: it is encoded in the factor
$\gamma^{-1}$ in the first term in the exponent. It is clear that
the momentum of perturbation labeled by ${\bf q}$ is actually equal to
\[
{\bf p} = (\gamma^{-1}q_{||},  {\bf q}^T) \; .
\]
Accordingly, the power spectrum (omitting the correction discussed
in section~\ref{h}) is given by
\[
\mathcal{P}_{\delta \theta}({\bf p})
\frac{d^3 p}{4\pi p^3} =
\frac{h^2}{16\pi^3}
\frac{\gamma d^3 p}{[(\gamma p_{||})^2 + ({\bf p}^T)^2]^{3/2}} \; .
\]
The same power spectrum charactersizes the adiabatic perturbations
in our scenario, and changing the notation ${\bf p} \to {\bf k}$
we obtain finally
\[
\mathcal{P}_\zeta ({\bf k}) = A_\zeta
\frac{p^3}{[(\gamma k_{||})^2 + ({\bf k}^T)^2]^{3/2}}
= A_\zeta \left(1 - \frac{3}{2} \frac{({\bf kv})^2}{k^2} \right)\; .
\]
Hence, we have arrived at the last term in \eqref{jul22-6}.
Again, neither the direction of ${\bf v}$ nor its length can be
unambiguously calculated because of cosmic variance; recall, however,
that the  value of $|{\bf v}|$, and hence the parameter
$c_2$ in  \eqref{jul22-6}, is logarithmically enhanced due to
the infrared effects, see \eqref{oct5-3}.

In the case of
the pseudo-Nambu-Goldstone mechanism of conversion of the phase
perturbations into the adiabatic ones, both terms
in the statistical ansotropy \eqref{jul22-6}
are small because of
the bound \eqref{jul27-10}. On the other hand, the effect
may be stronger in the case of the modulated decay mechanism.

\section{Conclusion}
\label{conclude}

We conclude by making a few remarks.

First, our
mechanism of the generation of the adiabatic perturbations
can work in any cosmological scenario that solves the
horizon problem of the hot Big Bang theory, including inflation,
bouncing/cyclic scenario, pre-Big Bang, etc. In some of these
scenarios (e.g., bouncing Universe),
the assumption that the phase perturbations
are superhorizon in conventional sense by the end of the
conformal rolling stage may be non-trivial. It would be of
interest to study also the opposite case, in which the phase
evolves for some time after the end of conformal rolling.

Second, we concentrated in this paper on the effect of
infrared radial modes, and employed the derivative expansion.
The expressions like \eqref{jul25-32}, which we obtained in this way,
must be used with caution, however. Bold
usage of  \eqref{jul25-32} would yield, e.g.,
non-vanishing equal-time commutator $[\theta ({\bf x}), \theta ({\bf
y})]$, which would obviously be a wrong result. The point is that the
formula  \eqref{jul25-32} is valid in the approximation ${\bf
v}=\mbox{const}$; with this understanding, the equal-time commutator
vanishes, as it should.

%\marginpar{\bf \tiny Paragraph removed}
%Third, a potentially observable effect we have studied in this
%paper is the statistical anisotropy. Unlike in the case of anisotropic
%inflation, the statistical anisotropy
%is a combination of a larger (order O($h$)) term which, however,
%statistical anisotropy
%decays as $k$ increases, and a smaller (order O($h^2 \log \Lambda$))
%term which is independent of $k$. We consider this feature as a
%potential smoking-gun property of our scenario.
%In view of strong
%cosmic variance at small $k$, the detection of this effect is
%probably difficult, unless the coupling $h$ is fairly large.

Finally, the
non-linearity of the field equation gives rise to the intrinsic
non-Gaussianity of the phase perturbations and, as a result,
adiabatic perturbations. The non-Gaussianity emerges at the order $O(h^2)$,
and may be sizeable
for large enough values of the coupling $h$.
The form of the non-Gaussianity is rather
peculiar in our scenario. Unlike in many other cases, the three-point
correlation function vanishes, while the four-point correlation
function of $\delta \theta$ (and hence of adiabatic perturbations)
involves the two-point correlator of the independent Gaussian field
$\delta \eta_*$. In view of the result of section~\ref{twoorders},
correlation functions of  $\delta \theta$ should be infrared-finite,
%\marginpar{\bf \tiny Changed}
at least at the order  $O(h^2)$.
This result shows also that the analysis
of the non-Gaussianity requires methods beyond the gradient expansion.
We leave this important issue for the future.

\acknowledgments
The authors are indebted to A.~Barvinsky, S.~Dubovsky,
A.~Frolov, D.~Gorbunov, E.~Komatsu, S.~Mironov,
V.~Mukhanov, S.~Mukohyama,
M.~Osipov, S.~Ramazanov and A.~Vikman for helpful discussions.
We are grateful to the organizers of the Yukawa International
Seminar ``Gravity and Cosmology 2010'', where part of this work has
been done, for hospitality.
%\marginpar{\bf Workshop}
 This work
has been supported in part by Russian Foundation for Basic
Research grant 08-02-00473, the Federal Agency for Sceince and
Innovations under state contract 02.740.11.0244 and the grant of
the President of Russian Federation NS-5525.2010.2.
The work of M.L. has been supported in part by Dynasty Foundation.

\end{document}